# Semi-analytical Framework for Modeling Strong Coupling of Quantum Emitters in Electromagnetic Resonators

Mohammad Abutoama[*], George Kountouris, Jesper Mørk, and Philip Trøst Kristensen

**ABSTRACT**

We present a semi-analytical framework for studying interactions between quantum emitters and general electromagnetic resonators. The method relies on the Lippmann-Schwinger equation to calculate the complex resonance frequencies of the coupled system based only on a single calculation for the electromagnetic resonator without the quantum emitter and with no fitting parameters. This is in stark contrast to standard approaches in the literature, in which the properties of the coupled system are fitted from calculated spectra. As an application example, we consider a recent dielectric cavity design featuring deep subwavelength confinement of light. We find the expected anti-crossing of the emitter and cavity resonance frequencies, and comparing to independent reference calculations, we find an extraordinary quantitative agreement with a relative error below one part in ten thousand. In order to unambiguously connect with the Jaynes-Cummings model, we derive an explicit expression relating the classical description of the emitter, as modeled by a spherical inclusion with a Lorentzian material response, to the dipole moment of the corresponding quantum optical model. The combined framework therefore enables classical calculations to be used for evaluating the coupling strength entering quantum optical theories in a transparent way.

## INTRODUCTION

Efficient light-matter interaction at the nanoscale is of high interest for many important applications such as single-photon sources [1-7] for quantum information and communication technology [8-11], which often rely on the effective interfacing of quantum emitters (QEs) with optical cavities. When a QE is placed in a nanostructured environment, such as an optical cavity for example, the light-matter interaction can be modified from that in free space. For relatively weak coupling, the spontaneous emission rate is found to be enhanced by the so-called Purcell effect [12]. In the strong coupling limit, on the other hand, the dynamics are characterized by coherent energy exchange between the QE and the cavity field, which can be observed in the frequency domain through a splitting of the spectrum into two peaks. This is illustrated in Fig. 1, which shows the absorption cross-section spectrum of a strongly coupled system consisting of a QE in the center of an optical cavity.

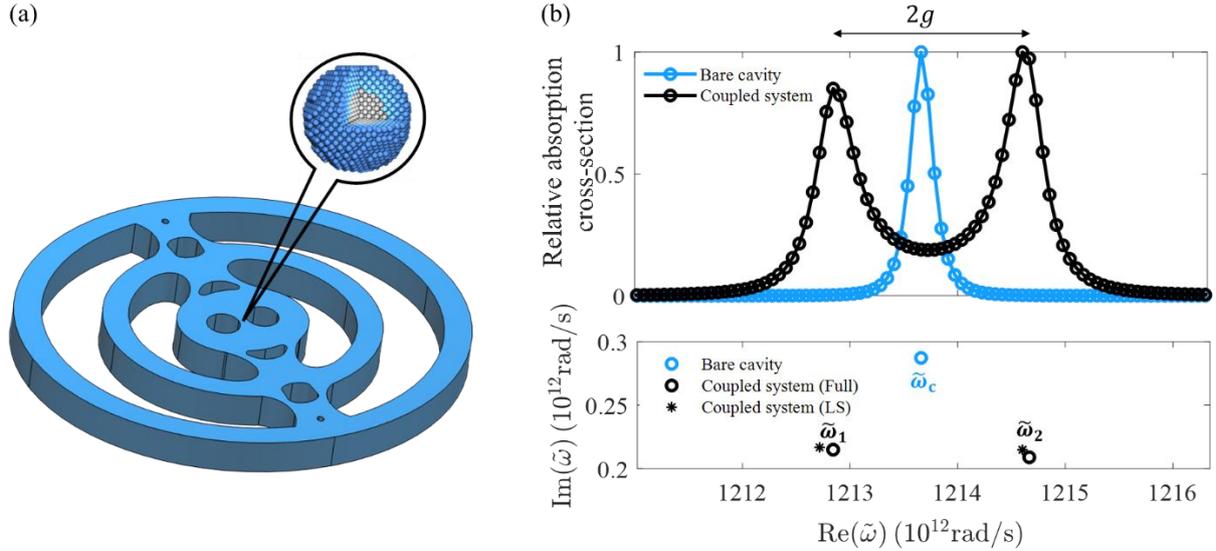

**Fig 1. (a)** General schematic of a QE-optical cavity hybrid system. **(b)** Top: Relative absorption cross-section of the bare cavity without (blue) and with (black) an embedded QE. The bare cavity spectrum features a single resonance peak, which corresponds to a single complex frequency $\tilde{\omega}_c$. The spectrum for the coupled system, on the other hand, shows two peaks, each corresponding to a complex frequency and for which the frequency splitting is proportional to the coupling strength $g$. Bottom: Complex frequency spectrum showing the discrete frequencies (circles) as well as the approximate values for the coupled system as obtained by the semi-analytical approach (stars).

Light-matter interaction in a QE-optical cavity hybrid system can be theoretically studied using different frameworks and under different approximations. In particular, a two-level system is known to behave like a Lorentz oscillator in the weak excitation limit [13] of linear response, and the use of a small volume of material with a Lorentzian frequency response is a popular method of modeling QE-optical cavity hybrid systems by mapping the problem onto the model system of two coupled harmonic oscillators [14-20]. In this physically appealing approach, the dynamics of the coupled system can be seen as arising from the coherent exchange of energy between the QE and the optical cavity, and the approach therefore effectively provides a connection between classical electrodynamics and the Jaynes-Cummings model of cavity quantum electrodynamics. Despite the considerable success and use of this approach, the mapping of the continuous classical electromagnetic field problem onto the discrete oscillators suffers from the lack of an underlying theory, which means that phenomenological fitting parameters in practice are required to account for the experimental results. To explicitly define the discrete oscillators and thereby alleviate the use of fitting parameters, we note that it is by now understood, that the general response of electromagnetic resonators can be conveniently and quite naturally described using the natural resonances of the electromagnetic system, the so-called quasi-normal modes (QNMs) [21-24] of the system, which are also known as resonant states [25-26]. From this point of view, the introduction of a Lorentz oscillator material within or near the electromagnetic resonator leads to the formation of an additional QNM [27-28], which is the origin of the additional peak in the spectrum, cf. Fig. 1. The coupling strength of the coupled system can thus be analyzed by investigating the complex frequency spectrum.

In this work, we theoretically examine the interaction between a single QE and an optical cavity and present a semi-analytical approach to calculate the coupling strength based only on the QNM of the bare cavity. The model relies on the Lippmann-Schwinger (LS) equation [29] to calculate the QNM frequencies of the coupled system with no fitting parameters. The proposed calculation scheme is valid for general electromagnetic resonators, be they dielectric, plasmonic, or hybrid. For the calculations in this work, we consider a particular optical cavity belonging to a relatively new class of dielectric cavities with deep subwavelength confinement, that are of contemporary interest both theoretically and experimentally [30-46]. This class of cavities, which we refer to as extreme dielectric confinement (EDC) cavities, opens intriguing possibilities for significantly enhancing light-matter interaction. Experimental results for a silicon EDC cavity designed using topology optimization were presented in Ref. [47], and a thorough numerical investigation using a simplified version of this design is available in Ref. [48]. Our calculations in this work are based on a slightly modified version of this design, as shown in Fig. 1.

The paper starts with the mathematical formulation of the scattering problem of interest. We begin with a brief description of the use of the LS equation for calculating the QNMs of the coupled system, after which we present a special case where we recover the result of the Jaynes-Cummings model and thereby provide an explicit expression connecting the dipole moment to the properties of the dispersive material forming the QE. In the following section, we provide an application example, in which we demonstrate the expected anti-crossing of the resonances when varying the QE-cavity detuning and compare the predictions of the semi-analytical approach to full electromagnetic reference calculations as well as to the predictions from the Jaynes-Cummings model. Summary and conclusions are given in the last section.

## RESULTS AND DISCUSSION

### Formulation

In this section, we describe the semi-analytical model for analyzing the scattering problem in Fig. 1(a) in which a QE is placed in the center of the optical cavity and thereby forms a strongly-coupled QE-optical cavity hybrid system. In particular, we will show that the QNM resonances of the coupled system, which describe the peaks of the spectrum in Fig. 1(b), can be calculated using the QNM of the bare optical cavity with no QE. Throughout the manuscript, we consider non-magnetic and piecewise isotropic materials for simplicity, but we note that the method generalizes immediately to more complicated material models.

The starting point is the LS equation, which provides the solution to a general scattering problem in terms of the background electric field Green function $\mathbf{G}_\text{B}(\mathbf{r}, \mathbf{r}', \omega)$ through an integral over all space of the form [24, 29]

$$\mathbf{E}_\text{tot}(\mathbf{r}, \omega) = \mathbf{E}_\text{in}(\mathbf{r}, \omega) + \frac{\omega^2}{c^2} \int \mathbf{G}_\text{B}(\mathbf{r}, \mathbf{r}', \omega) \Delta\varepsilon(\mathbf{r}', \omega) \mathbf{E}_\text{tot}(\mathbf{r}', \omega) dV', \qquad (1)$$

in which $\mathbf{E}_\text{tot}(\mathbf{r}, \omega)$ denotes the total electric field, $\mathbf{E}_\text{in}(\mathbf{r}, \omega)$ is the incoming electric field, c is the speed of light in vacuum, and $\Delta\varepsilon(\mathbf{r}, \omega) = \varepsilon_\text{R}(\mathbf{r}, \omega) - \varepsilon_\text{B}(\mathbf{r}, \omega)$ denotes the difference between the total permittivity distribution and the background permittivity $\varepsilon_\text{B}(\mathbf{r}, \omega)$, which in general is a function of position $\mathbf{r}$ and angular frequency $\omega$.

Importantly, the incoming electric field is a solution to the wave equation in a geometry described by the background permittivity, so that all scattering is caused by the change in permittivity $\Delta\varepsilon(\mathbf{r}, \omega)$. Similarly, the background Green function is the Green function for the geometry described by the background permittivity. In typical scattering calculations, one will often choose the background permittivity to be constant, in which case the background Green function is known analytically, and one can use combinations of plane waves for the incoming field.

The QNMs are solutions to the sourceless wave equation subject to a suitable radiation condition, such as the Silver-Müller radiation condition in the case of homogeneous media [24]. Therefore, we can calculate the QNMs corresponding to a given geometry defined by $\Delta\varepsilon(\mathbf{r},\omega)$ as the solutions to the LS equation with no incoming field [49-50],

$$\tilde{\mathbf{f}}_n(\mathbf{r}) = \frac{\tilde{\omega}_n^2}{c^2} \int \mathbf{G_B}(\mathbf{r},\mathbf{r}',\tilde{\omega}_n)\Delta\varepsilon(\mathbf{r}',\tilde{\omega}_n)\tilde{\mathbf{f}}_n(\mathbf{r}')dV', \qquad (2)$$

where $\tilde{\mathbf{f}}_n(\mathbf{r})$ denotes the *n*'th electric field QNM with corresponding complex frequency $\tilde{\omega}_n = \omega_n - i\gamma_n$, in which the imaginary part describes the dissipation, and the associated quality factor can be calculated as $Q_n = \omega_n/2\gamma_n$. In particular, the QNM of the bare cavity in Fig. 1 above can be calculated using Eq. (1) by choosing $\mathbf{G_B}(\mathbf{r},\mathbf{r}',\omega)$ to be the Green function of free space and choosing $\varepsilon_R(\mathbf{r},\omega)$ to be the permittivity distribution defining the optical cavity. Similarly, the QNMs of the coupled QE-optical cavity system can be calculated by applying the same equation but choosing $\varepsilon_R(\mathbf{r},\omega)$ to be the permittivity of the optical cavity including the embedded QE. A third option for calculating the QNMs of the coupled system - which is the one we shall exploit in this work - is to choose the Green function to be that of the bare cavity and $\Delta\varepsilon(\mathbf{r},\omega)$ to be the permittivity change due to the QE only. This choice is illustrated in Figure 2.

Following Refs. [15, 16, 19], we model the QE as a small sphere of homogeneous but dispersive permittivity in the form of a single Lorentzian response,

$$\varepsilon_{\mathrm{QE}}(\omega) = \varepsilon_\infty + \frac{f\omega_{\mathrm{QE}}^2}{\omega_{\mathrm{QE}}^2 - \omega^2 - 2i\gamma_{\mathrm{QE}}\omega}, \qquad (3)$$

where $\varepsilon_\infty$ is a constant background permittivity, $f$ is the oscillator strength of the electronic transition in the material, and $\gamma_{\mathrm{QE}}$ and $\omega_{\mathrm{QE}}$ denote, respectively, the damping rate and the resonance angular frequency of this transition. Since we will be investigating the case where the QE is placed inside the material at the cavity center, we take the constant background permittivity to be that of the high-index cavity material. The permittivity perturbation that enters the LS equation is then given by: $\Delta\varepsilon_{\mathrm{QE}}(\omega) = \frac{f\omega_{\mathrm{QE}}^2}{\omega_{\mathrm{QE}}^2 - \omega^2 - 2i\gamma_{\mathrm{QE}}\omega}$.

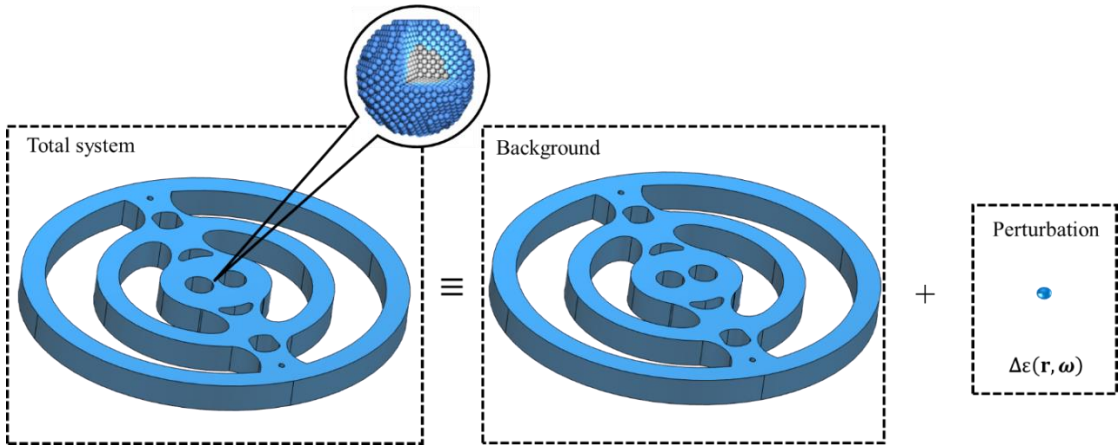

**Fig 2.** The general scheme of the calculation approach: the geometry of the coupled QE-optical cavity system can be divided into the bare cavity in free space and an added permittivity perturbation $\Delta\varepsilon(\mathbf{r}, \omega)$ due to the QE.

Equation (2) is a Fredholm integral equation of the second kind in which the electric field QNM $\tilde{\mathbf{f}}_n(\mathbf{r})$ appears on both sides of the equality. The general solution calls for numerical calculation schemes, but for the present problem we can turn it into a simple algebraic equation by means of a few simplifying assumptions. First, we make use of the fact that the electromagnetic response of the bare cavity is well represented by a single QNM for frequencies close to the resonance and for positions close to the cavity center, as detailed in Ref. [48]. Therefore, we can approximate the Green tensor by use of this QNM only as

$$\mathbf{G}_B(\mathbf{r}, \mathbf{r}', \omega) \approx \frac{c^2}{2\omega} \frac{\tilde{\mathbf{f}}_c(\mathbf{r})\tilde{\mathbf{f}}_c(\mathbf{r}')}{\tilde{\omega}_c - \omega} \tag{4}$$

where $\tilde{\mathbf{f}}_c(\mathbf{r})$ is the normalized electric field of the QNM of interest for the bare cavity at the position $\mathbf{r}$, and $\tilde{\omega}_c = \omega_c - i\gamma_c$ is the corresponding complex resonance frequency. Next, inserting the single-QNM approximation of the Green tensor in Eq. (2) and carrying out the integration assuming the field to be approximately constant across the volume of the QE, we can rewrite Eq. (2) as

$$\tilde{\mathbf{f}}_n(\mathbf{r}_{QE}) \left[ 1 - \frac{\tilde{\omega}_n}{2} \frac{\tilde{\mathbf{f}}_c(\mathbf{r}_{QE}) \cdot \tilde{\mathbf{f}}_c(\mathbf{r}_{QE})}{\tilde{\omega}_c - \tilde{\omega}_n} \Delta\varepsilon_{QE}(\tilde{\omega}_n) V_{QE} \right] \approx 0, \tag{5}$$

In which $\mathbf{r}_{QE}$ and $V_{QE}$ denote the position and the volume of the QE, respectively. For non-trivial solutions, we demand that

$$1 - \frac{\tilde{\omega}_n}{2} \frac{\tilde{\mathbf{f}}_c(\mathbf{r}_{QE}) \cdot \tilde{\mathbf{f}}_c(\mathbf{r}_{QE})}{\tilde{\omega}_c - \tilde{\omega}_n} \frac{f\omega_{QE}^2}{\omega_{QE}^2 - \tilde{\omega}_n^2 - 2i\gamma_{QE}\tilde{\omega}_n} V_{QE} = 0. \tag{6}$$

The left hand side of this equation defines a complex function for which the QNM frequencies $\tilde{\omega}_n$ appear as zeros. Noting that the denominator in the permittivity function derives from the two approximate resonances at $\tilde{\omega}_{QE} = \omega_{QE} - i\gamma_{QE}$ and $-\tilde{\omega}_{QE}^* = -\omega_{QE} - i\gamma_{QE}$, we can rewrite this expression in the approximate form

$$1 - \frac{\tilde{\omega}_n}{2} \frac{\tilde{\mathbf{f}}_c(\mathbf{r}_{QE}) \cdot \tilde{\mathbf{f}}_c(\mathbf{r}_{QE})}{\tilde{\omega}_c - \tilde{\omega}_n} \frac{f\omega_{QE}^2}{(\tilde{\omega}_{QE} - \tilde{\omega}_n)(\tilde{\omega}_{QE}^* + \tilde{\omega}_n)} V_{QE} = 0 \tag{7}$$

from which we expect two resonances in the vicinity of $\tilde{\omega}_n \approx \tilde{\omega}_c \approx \tilde{\omega}_{QE}$. With this ansatz, we approximate the non-resonant factor as $\tilde{\omega}_n/(\tilde{\omega}_{QE}^* + \tilde{\omega}_n) \approx \frac{1}{2}$, and we can then rewrite the equation as

$$(\tilde{\omega}_c - \tilde{\omega}_n)(\tilde{\omega}_{QE} - \tilde{\omega}_n) - g^2 = 0 \tag{8}$$

where we have defined $g^2 = fV_{QE}\omega_{QE}^2/(4\varepsilon_R(\mathbf{r}_{QE})v_c)$. In this expression, $\varepsilon_R(\mathbf{r})$ is the dispersionless permittivity distribution of the bare cavity, and $v_c$ is the generalized effective mode volume [49],

$$v_c = \frac{\langle\langle \tilde{\mathbf{f}}_c | \tilde{\mathbf{f}}_c \rangle\rangle}{\varepsilon_R(\mathbf{r}_{QE})\tilde{\mathbf{f}}_c(\mathbf{r}_{QE}) \cdot \tilde{\mathbf{f}}_c(\mathbf{r}_{QE})} \tag{9}$$

in which $\langle\langle \tilde{\mathbf{f}}_c | \tilde{\mathbf{f}}_c \rangle\rangle$ denotes the QNM normalization [51-54]. Solving Eq. (8), we can express the complex eigenfrequencies of the coupled system as

$$\widetilde{\omega}_{1,2} = \frac{\omega_c + \omega_{QE}}{2} - i\frac{\gamma_c + \gamma_{QE}}{2} \pm \sqrt{g^2 + \frac{1}{4}[(\omega_c - \omega_{QE}) - i(\gamma_c - \gamma_{QE})]^2} \qquad (10)$$

**Connection to the Jaynes-Cummings model**

The Jaynes-Cummings (JC) model [55] describes the interaction of an optical cavity and a QE without losses in second quantization. The coupled system is described by the Hamiltonian

$$H = \hbar\omega_c a^\dagger a + \hbar\omega_{QE}\sigma^\dagger\sigma + \hbar(ga^\dagger\sigma + g^*\sigma^\dagger a) \qquad (11)$$

in which $\hbar$ is the reduced Planck constant, and the operators $a$ ($a^\dagger$) and $\sigma$ ($\sigma^\dagger$) are lowering (raising) operators of the cavity field and the QE, respectively. The cavity field inherently behaves as a harmonic oscillator [55], whereas the QE can be well represented as a two level system. We note, however, that for the calculations in this work, in which we focus on the single-excitation subspace, we can take the QE to be a harmonic oscillator as well; this also justifies the alternative analysis in the appendix. The parameter $g$ denotes the coupling strength, which is connected to the dipole moment of the QE as [56]

$$\hbar g = \sqrt{\frac{\hbar\omega_c}{2\varepsilon_0}}\mu_{QE} \cdot \tilde{\mathbf{f}}_c(\mathbf{r}_{QE}) \qquad (12)$$

where $\varepsilon_0$ is the permittivity of free space. The operators evolve in time according to the Heisenberg equations of motion. In order to account for dissipation of energy, we follow the general ideas of Refs. [56-57] and add imaginary parts to the frequencies along with fluctuating source terms. With this approach, we find that the system dynamics can be written in the form

$$\partial_t \begin{bmatrix} a \\ \sigma \end{bmatrix} = \begin{bmatrix} -i\widetilde{\omega}_c & -ig \\ -ig & -i\widetilde{\omega}_{QE} \end{bmatrix}\begin{bmatrix} a \\ \sigma \end{bmatrix} + \begin{bmatrix} F_c \\ F_{QE} \end{bmatrix} \qquad (13)$$

where $F_c$ and $F_{QE}$ are fluctuating white noise terms connected with the dissipation, which are important for preserving the commutation relations for the system operators in time [56-57]. From Eq. (13), it is clear that the system dynamics are governed by the eigenvalues of the matrix, and by direct calculation, we find the complex frequencies of the coupled system in the exact form of Eq. (10).

By comparing the results of this fully quantum mechanical approach in the limit $\widetilde{\omega}_c = \widetilde{\omega}_{QE}$ to the solution of the semi-analytical approach based on the LS equation, we can now directly connect the dipole moment in the Jaynes-Cummings model to the oscillator strength and volume of the dispersive material making up the QE. Alternatively, we can follow the approach of Ref [58] and calculate the solution to the quantum mechanical problem in terms of a LS equation, as we do in the appendix. In both cases, we find that

$$\mu_{QE} = \sqrt{\frac{f\hbar\varepsilon_0 V_{QE}\omega_{QE}}{2}} \qquad (14)$$

**Application example**

In this section, we provide an example application of the theory. We consider the optical cavity in Figs. 1 and 2, which is derived from the design in Ref. [48] by slightly modifying it to use a relative permittivity of $\varepsilon_{InP} = 10.02$ corresponding to indium phosphide around the target wavelength $\lambda_0 = 1550$ nm [59]. For the QE, we consider a small sphere with radius 20 nm and placed at the center of the cavity. The material of the QE is described by the dispersive permittivity in Eq. (3) with oscillator strength $f=7 \times 10^{-3}$ and $\widetilde{\omega}_{QE} = \widetilde{\omega}_c$ to perfectly match the bare cavity resonance. As an experimentally relevant way of probing the resonances in the system, we first calculate the absorption cross section of the system with and without QE, as detailed in the appendix. The results are shown in the top panel of Fig. 1(b) and show the characteristic splitting of a single resonance into a double-peaked spectrum when the QE is included.

From the discussion above, we expect the single peak in the spectrum of Fig. 1(b) to be attributable to a single QNM frequency. Instead of the formulation in Eq. (2), we calculate the QNM fields numerically with the finite element method, as detailed in the appendix. We find that the bare optical cavity supports a QNM with an angular resonance frequency of $\widetilde{\omega}_c = (1213.66 - i0.287) \, 10^{12} \text{rad s}^{-1}$ corresponding to a $Q = 2114$, as shown by the blue open circle in the bottom panel of Fig. 1(b). The mode profile of this QNM is shown in Fig. 3(a) and it has a generalized effective mode volume of $v_c = (0.691 + i0.002) \, (\lambda_0/2n)^3$.

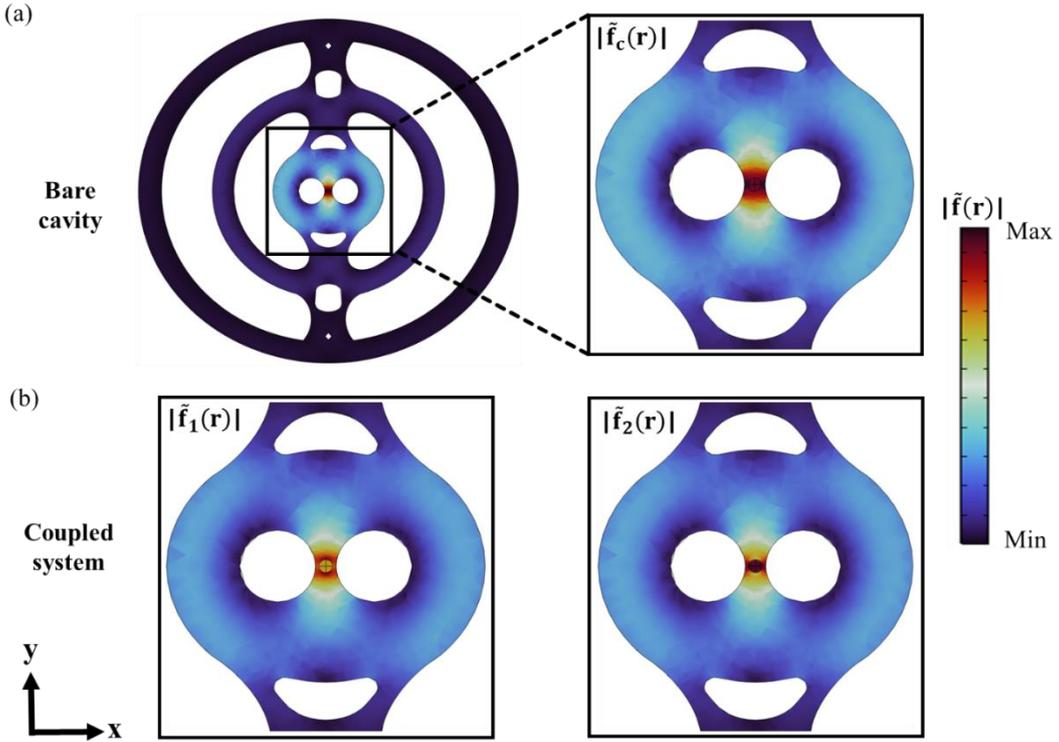

**Fig 3. (a)** Mode profile showing the magnitude of the electric field QNM of interest in the bare cavity. The field is strongly localized at the cavity center. **(b)** Mode profiles of the two QNMs of interest in the coupled QE-optical cavity hybrid system.

Next, we consider the effect of including the QE. By direct numerical calculations, we find the two complex frequencies $\widetilde{\omega}_1 = (1212.84 - i0.215) \, 10^{12} \text{rad s}^{-1}$ and $\widetilde{\omega}_2 = (1214.67 - i0.209) \, 10^{12} \text{rad s}^{-1}$ as shown by the black open circles in the bottom part of Fig. 1(b). The mode profiles of the two corresponding QNMs are shown in Fig. 3(b). We can now compare the results of the full numerical calculation to the approximate resonance frequencies resulting from the LS equation. To this end, we solve Eq. (6) iteratively by numerical means and without the assumption $\widetilde{\omega}_n \approx \widetilde{\omega}_c \approx \widetilde{\omega}_{QE}$, and we find the approximate frequencies $\widetilde{\omega}_1^{LS} = (1212.72 - i0.217)\,10^{12}\text{rad s}^{-1}$ and $\widetilde{\omega}_2^{LS} = (1214.60 - i0.215)\,10^{12}\text{rad s}^{-1}$ as shown by black stars in the bottom part of Fig. 1(b). The corresponding relative errors are on the order of one part in ten thousand for the real parts and one percent for the imaginary part. The semi analytical approach evidently works remarkably well for predicting the coupling in this QE-optical cavity hybrid system.

We note that these calculations were performed with the mesh resulting from two refinements of the original mesh, as discussed in the appendix. Based on the convergence study, we expect the error on the real and imaginary parts of the stated numbers to be on the order of 1 and 0.01, respectively, as calculated by comparing to the best estimate of the true value for the case of the bare cavity. Notably, this accuracy does not justify the stated number of digits. Nevertheless, we include the additional digits to highlight the fact that since all calculations were done on the same mesh, the results are internally consistent to a higher accuracy than the estimate based on the absolute error. Essentially, the numerical error stemming from discretization and the finite size of the calculation domain affects all the calculations in a similar way.

To further examine the behavior of the coupled system, we consider the change in the spectrum when detuning the QE resonance with respect to the bare cavity resonance. The results are summarized in Fig. 4, which shows the real parts of both resonance frequencies in the coupled QE-optical cavity hybrid system as calculated with the full numerical simulations as well as the approximate LS equation. Comparing the two datasets, we find relative errors smaller than one part in ten thousand over the full spectral range of interest. In addition, we show the results from Eq. (10), which are identical to the prediction of the Jaynes-Cummings model with the appropriate scaling of the dipole moment from Eqs. (12) and (14). In this case, we find relative errors close to one part in a thousand.

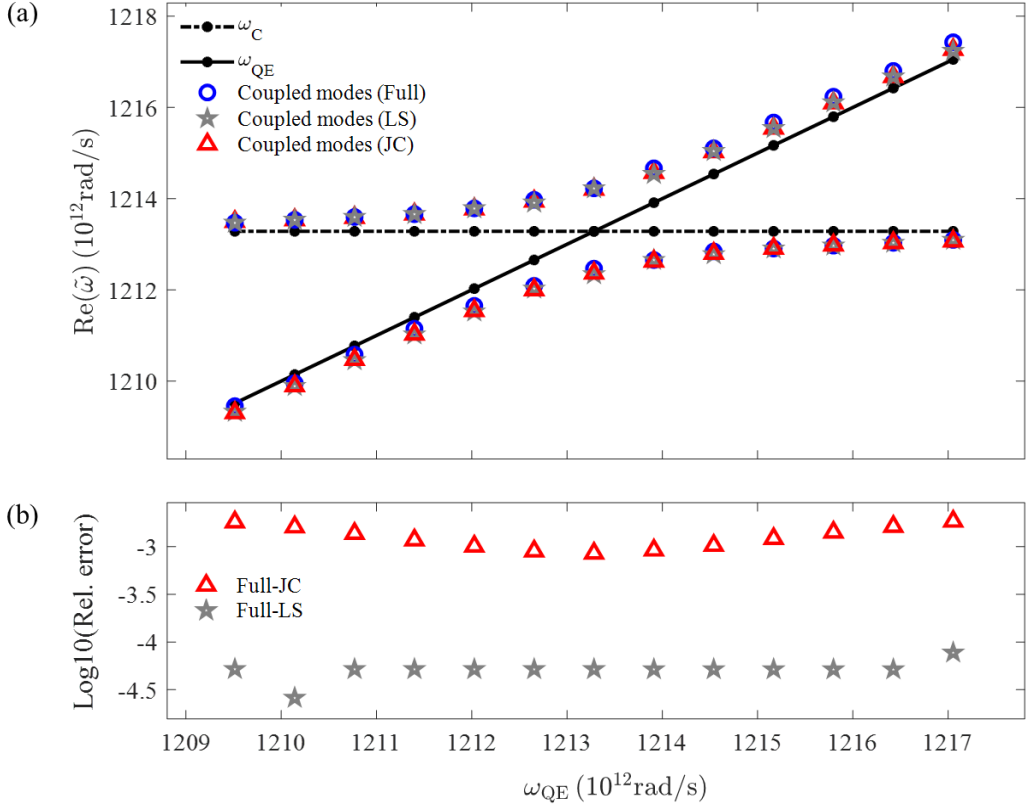

**Fig 4. (a)** Eigenfrequencies of the coupled QE-optical cavity hybrid system when detuning the QE resonance with respect to the cavity frequency, showing the expected anti-crossing of the frequencies. Blue circles, gray stars, and red triangles correspond to the full numerical solution (Full), the semi-analytical approach (LS) and the Jaynes-Cummings model, respectively. The dashed horizontal black line shows the resonance frequency of the optical cavity, while the solid black line indicates the angular frequency of the QE. **(b)** Corresponding relative errors of the semi-analytical approach (Full-LS) and the Jaynes-Cummings model (Full-JC) when compared t the full numerical solution.

The quantitative agreement between the LS equation and the reference calculations underscores the usefulness of the semi-analytical approach for precise predictions of the resonance frequencies of the QE-optical cavity hybrid system.

## SUMMARY AND CONCLUSIONS

We theoretically investigated the interaction between a single QE – as modeled by a small volume of material with a Lorentzian response – and a general electromagnetic resonator – such as an optical cavity, a plasmonic particle, or a combination – and presented a semi-analytical approach to calculate the coupling of the two into a hybrid system. The approach relies on the LS equation to calculate the resonance frequencies of the coupled system without the need for fitting to calculated spectra. Once the complex frequency and generalized effective mode volume of the QNM of interest in the bare cavity is calculated – typically by numerical means – the complex frequencies of the coupled system can be readily calculated to high accuracy based only on the properties of the QE. As a special case, we recover the result of the Jaynes-Cummings model and thereby provide an explicit expression connecting the dipole moment to the oscillator strength and volume of the dispersive material making up the QE in this model. As an example application, we investigated the coupling of a QE to a dielectric nanocavity featuring deep subwavelength confinement. By detuning the QE resonance frequency we found the expected anti-crossing of the QE and cavity resonance frequencies, and by comparing to full numerical reference calculations, we found relative errors smaller than one part in ten thousand over the full frequency range of interest.

Efficient and transparent tools for modeling light-matter interaction in coupled systems at the nanoscale are important from a theoretical point of view as well as for applications in quantum technology. Based on the extraordinary quantitative agreement with the reference calculations, the semi-analytical approach presented in this article provides one such tool, which enables classical calculations to be used for evaluating the coupling strength entering quantum optical theories in a precise and transparent way.

## ASSOCIATED CONTENT

**Supporting Information**
The Supporting Information is available free of charge at http://pubs.acs.org.
Cavity design and numerical calculations, QNM convergence study and explicit expression of the coupling strength by connecting the classical oscillator strength of the emitter to the dipole moment of the corresponding quantum optical model.

## AUTHOR INFORMATION

**Corresponding Author**

**Mohammad Abutoama** − *DTU Electro, Technical University of Denmark, Ørsteds Plads, building 343, 2800 Kgs. Lyngby, Denmark; NanoPhoton - Center for Nanonphotonics, Ørsteds Plads, building 345A, 2800 Kgs. Lyngby, Denmark; orcid.org/0000-0002-4286-0434; Email: moabo@fotonik.dtu.dk*

**George Kountouris** − *DTU Electro, Technical University of Denmark, Ørsteds Plads, building 343, 2800 Kgs. Lyngby, Denmark; NanoPhoton - Center for Nanonphotonics, Ørsteds Plads, building 345A, 2800 Kgs. Lyngby, Denmark; orcid.org/0000-0003-4750-8701; Email: gkoun@dtu.dk*

**Jesper Mørk** − *DTU Electro, Technical University of Denmark, Ørsteds Plads, building 343, 2800 Kgs. Lyngby, Denmark; NanoPhoton - Center for Nanonphotonics, Ørsteds Plads, building 345A, 2800 Kgs. Lyngby, Denmark; Email: jesm@dtu.dk*

**Philip Trøst Kristensen** − *DTU Electro, Technical University of Denmark, Ørsteds Plads, building 343, 2800 Kgs. Lyngby, Denmark; NanoPhoton - Center for Nanonphotonics, Ørsteds Plads, building 345A, 2800 Kgs. Lyngby, Denmark; orcid.org/0000-0001-5804-1989; Email: ptkr@dtu.dk*

**Author Contributions**

The manuscript was written through contributions of all authors. All authors have given approval to the final version of the manuscript.

**Funding/ ACKNOWLEDGMENT**

This work was supported by the Danmark Grundforskningsfond (DNRF147); Danmarks Frie Forskningsfond (0164-00014B). M.A. was partially supported by the Planning and budgeting committee of the Israeli council for higher education (VATAT).

**Notes**

The authors declare no competing financial interest.

**Supporting Information**

**Cavity design and numerical calculations**

The specific cavity is designed to be manufactured from a 240 nm thick membrane of material with refractive index n=3.165 embedded in air, and the dimensions of the cavity are shown in Fig. S1 below. As noted in the main text, it supports a cavity mode with a wavelength close to the target wavelength of $\lambda_0 = 1550$ nm.

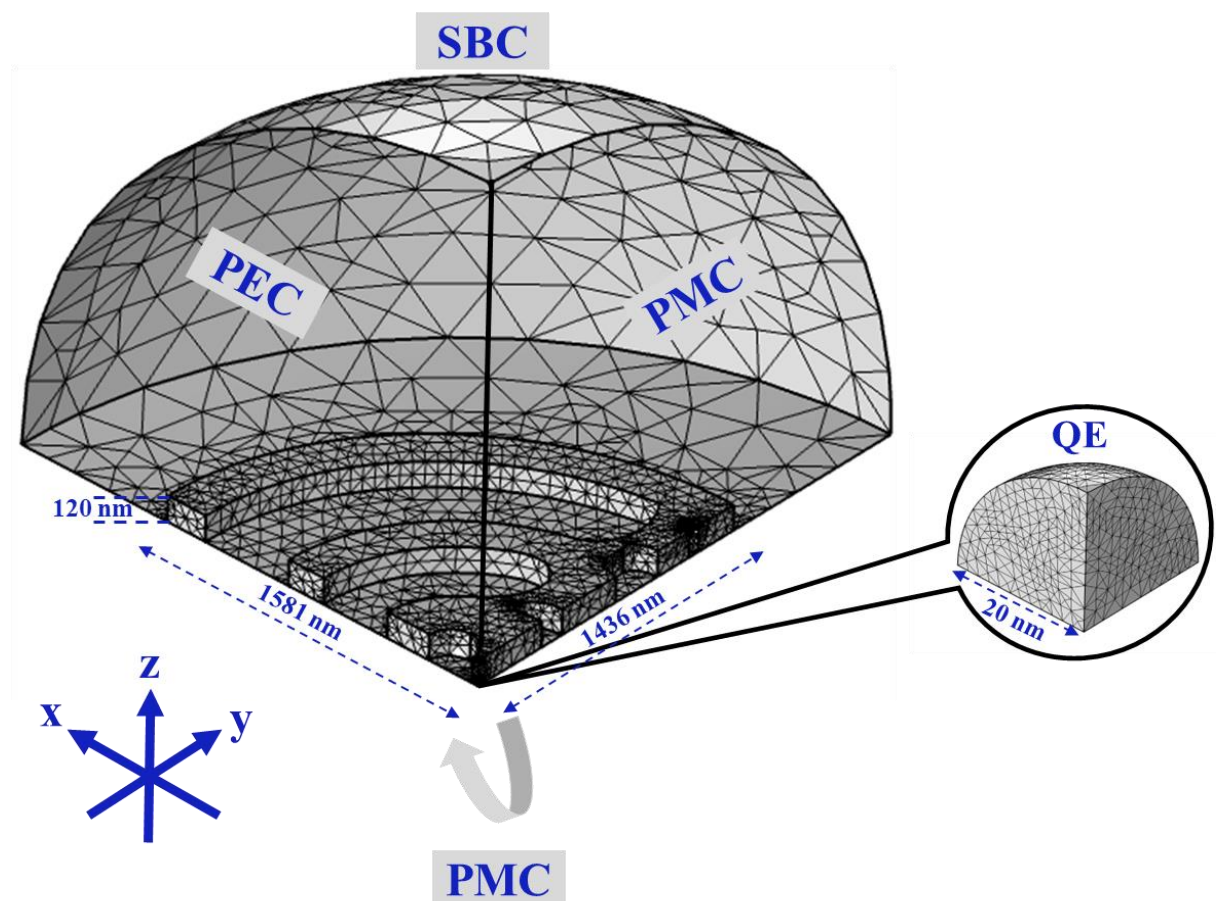

Fig. S1. Octant model for the QNM calculation exploiting the cavity mode symmetry. PEC: perfect electric conductor, PMC: perfect magnetic conductor, SBC: scattering boundary condition.

All numerical solutions of Maxwell's equations were performed with the finite element method as implemented in Comsol Multiphysics, and in all calculations, the calculation domain was truncated by use of a first-order scattering boundary condition (SBC). For the convergence study, a number of different meshes were created by successive refinement, as detailed below. The stated QNM frequencies as well as the absorption cross section calculations in the main text were calculated with a full domain size with radius $R=1.5\lambda_0$ and a mesh resulting from two refinements. Although we could go to a finer mesh for the QNM calculations, this was not feasible for the absorption cross sections. The use of the same mesh ensures that the results of the two calculations are consistent, since the residual error from the finite mesh and calculation domain size influence both calculations in a similar way. When calculating the QNMs of the bare cavity, we turned off the QE by setting its oscillator strength to 0 so as to use the exact same mesh.

The absorption cross-sections were calculated through an integral over the QE volume as $\sigma_{abs} = (1/I_0) \int_{V_{QE}} P(\mathbf{r},\omega) \, dv$, where $I_0$ is the intensity of the incident light, and P is the power loss density in the QE given by Poynting's theorem as [1]: $P(\mathbf{r},\omega) = \mathbf{J}(\mathbf{r},\omega) \cdot \mathbf{E}(\mathbf{r},\omega)$, in which **J** is the current density and **E** is the local electric field in the emitter. The current density can be expressed in terms of the electric field as

$$\mathbf{J}(\mathbf{r},\omega) = i\omega\varepsilon_0 \frac{f\omega_{QE}^2}{\omega_{QE}^2 - \omega^2 - 2i\gamma_{QE}\omega} \mathbf{E}(\mathbf{r},\omega) \tag{A1}$$

We consider this measure to be a convenient and experimentally relevant approach for probing the spectral response of the system. With the chosen material parameters, we changed the system between the two cases of interest by changing the oscillator strength between $f = 0$ and $f = 7 \times 10^{-3}$. Setting $f = 0$, however, leads to a vanishing absorption cross section, so in practical calculations of the bare cavity, for which the material is assumed to be non-absorbing in the frequency range of interest, we considered instead a very small oscillator strength of $f = 7 \times 10^{-9}$. In this way, the relative absorption cross section provides the relevant spectral features of the electromagnetic response of the bare cavity. The results are shown in the top part of Fig. 1(b) in the main text where the blue curve features a single peak at $\omega = 1213.66 \; 10^{12} \mathrm{rad \, s^{-1}}$, which matches the real part of the QNM resonance frequency of the bare cavity. Similarly, the black curve shows two peaks at $\omega = 1212.84 \; 10^{12} \mathrm{rad \, s^{-1}}$ and $\omega = 1214.60 \; 10^{12} \mathrm{rad \, s^{-1}}$, respectively, which also match the two QNM resonance frequencies of the QE-optical cavity hybrid system.

**QNM convergence study**

For a coordinate system as shown in Fig. S1, the QNM field of the cavity is symmetric with respect to the *xy*- and *yz*-planes, while it is antisymmetric with respect to the *xz*-plane [2]. Exploiting this symmetry, we used perfect electric conductor (PEC) and perfect magnetic conductor (PMC) boundary conditions to reduce the calculation domain to one-eighth of the original size for the convergence study in this section, which significantly reduced the computational time.

Following [2-3], a convergence study was performed by varying the domain size and mesh discretization. The investigation was carried out for 5 different mesh discretizations and for 12 domain sizes with radii ranging from *R*=1.5λ₀ to *R*=2.6λ₀ in steps of 0.1λ₀. In the case of the finest mesh, the calculations were performed for the range of radii *R*=1.5λ₀ to *R*=2.1λ₀ in steps of 0.1λ₀ due to the extended calculation times. The five curves in Fig. S2. (a) show the variation of the calculated eigenfrequencies in the complex frequency plane for the different discretizations used in the simulations. The black curve corresponds to the initial coarsest mesh, while the purple curve corresponds to the finest mesh. Each refinement of the mesh was performed by splitting the elements of the previous discretization, and we characterize each mesh by the average element size, as calculated by averaging the longest side *h* of each mesh element in the entire domain [2]. The individual data points comprising each curve are the complex eigenfrequencies as calculated for different domain sizes but with the same mesh discretization. The curves form an inwards spiral around a central point, which we take to be the nominal correct value for the case of an infinite calculation domain [2]. For relatively coarse meshes, the spirals are distorted due to additional numerical errors stemming from the finite mesh size. The red circle inside each spiral shows the average of the data points forming the spiral, and we use this average as an estimate of the nominal correct value.

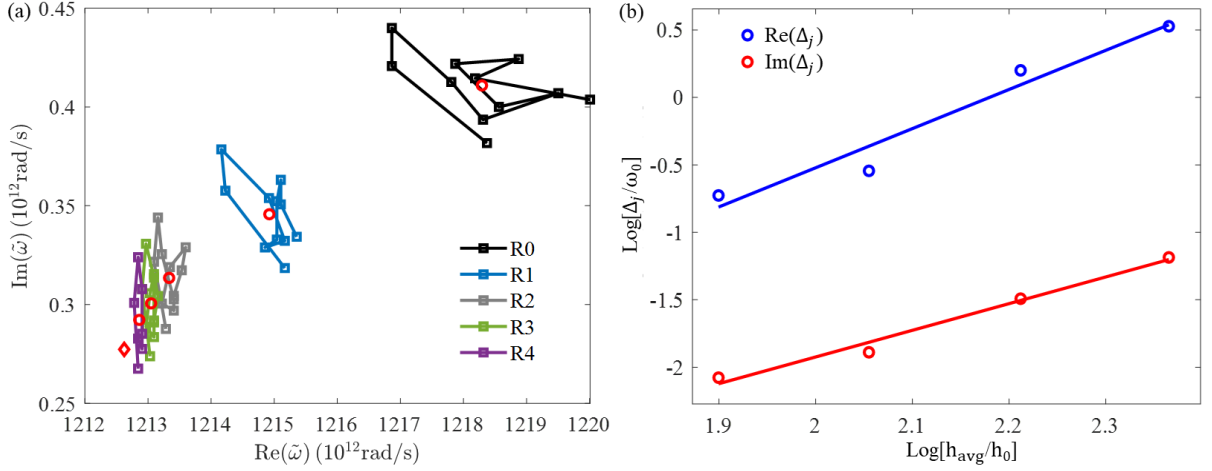

**Fig. S2. (a).** Numerical oscillations of the complex eigenfrequency for different sizes of the domain and mesh elements. The black spiral (R0) corresponds to the coarsest mesh, while the purple (R4) corresponds to the finest mesh used in the simulations. The data points forming each spiral are the complex eigenfrequencies for different domain sizes. The red circles indicate the average complex eigenfrequency for each data set. **(b).** Logarithm of the difference $\Delta_j = \widetilde{\omega}_{j+1} - \widetilde{\omega}_j$ between the average complex eigenfrequencies corresponding to two consecutive discretizations. Blue and red data points correspond to the real and the imaginary parts of the difference, respectively. The average is scaled by $\omega_0 = 10^{12}$ rad s$^{-1}$ and shown as a function of the logarithm of the average element size $h$ in units of $h_0=1$ nm. The red diamond in Fig. S2 (a) indicates the estimate of the true value of the resonance frequency corresponding to the underlying continuous problem as estimated through the linear fit in Fig. S2 (b).

To investigate the convergence of the complex eigenfrequency in a more quantitative way, Fig. S2 (b) shows the logarithm of the real (blue) and imaginary (red) parts of the difference in the calculated complex eigenfrequencies between two consecutive discretizations as a function of the logarithm of the average element size [2]. The points fall approximately on straight lines in this double-logarithmic plot, which indicates the expected polynomial convergence with mesh element size. Assuming the error to be polynomial, we can estimate the true value of the complex eigenfrequency of the underlying continuous problem corresponding to the limit of vanishing mesh size and infinite calculation domain [3]. With this approach, we find $\widetilde{\omega}_c = (1212.6(2) - i0.27(1)) \, 10^{12}$ rad s$^{-1}$, as indicated by the red diamond in Fig. S2 (a). As a conservative estimate of the error on this number, we take the difference to the best direct calculation with four refinements. In this way, we find the estimated error to be 0.2 and 0.01 for the real and the imaginary part, respectively.

For two different discretizations, we now consider the complex eigenfrequencies of the coupled system, as predicted by the LS equation, and compare it to the results of full numerical calculations. The results are shown in Fig. S3. The blue and red solid spirals show the numerical oscillations of the QNM frequency for the bare cavity for the second (R2) as well as the fourth (R4) mesh refinement. For each calculation, we can set up and solve Eq. (6) in order to find the corresponding approximation to the complex frequencies of the QE-optical cavity hybrid system. These results are indicated by the black dashed curves for the R4 mesh refinement. To assess the accuracy of the calculation, we also directly calculated the results for the coupled system using the same mesh, by introducing the permittivity describing the QE. These results are shown by the blue and red dashed curves in Fig. S3.

As is evident from these figures, the QNM frequencies of the coupled QE-optical cavity hybrid system change in a very similar manner to that of the bare cavity when increasing the calculation domain size. The black circle inside each spiral indicates the average of the complex QNM frequencies forming the spiral.

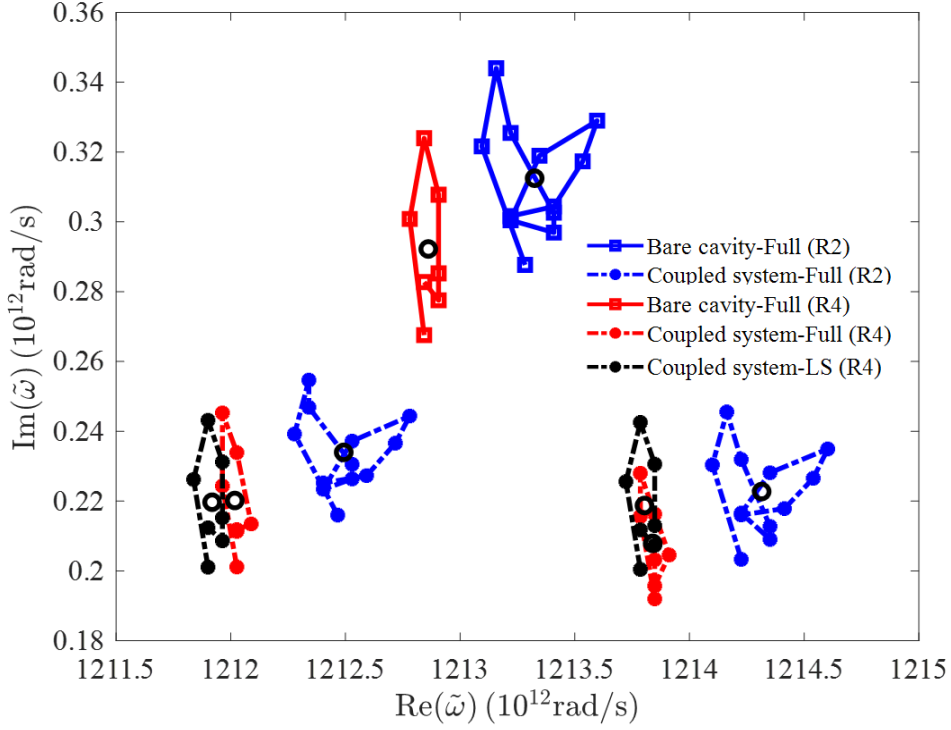

**Fig. S3.** Numerical oscillations of the QNM frequencies in the complex plane for varying calculation domain sizes. The blue and red solid spirals correspond to the second (R2) and the fourth (R4) mesh refinement of the bare cavity, respectively, and the corresponding dashed spirals show the variation in the QNM frequencies of the coupled QE-optical cavity hybrid system. The black spirals correspond to the complex eigenfrequencies calculated using the LS approximation equation based on the data for the bare cavity as calculated with the fourth mesh refinement. In all cases, the black circles inside each spiral indicate the average of the frequencies forming the spiral.

**Explicit expression of the coupling strength $g$ by connecting the classical oscillator strength of the emitter to the dipole moment of the corresponding quantum optical model**

As an alternative to the derivations presented in the main text, we can also connect the classical electrodynamics model to the quantum optical model by treating both in the framework of a scattering problem, for which the solution can be written by use of the LS equation. Specifically, we consider the problem of an incoming electric field $\mathbf{E}_{in}(\mathbf{r}, \omega)$, which is assumed to be a solution to the Maxwell wave equation in the background system depicted in Fig. 2 in the main text. Notably, the background system consists of the bare cavity embedded in air.

For the classical electrodynamics problem, the solution is given by Eq. (1) of the main text. At the QE center $\mathbf{r}'$, and assuming the total electric field to be approximately constant throughout the volume of the QE, we find that

$$\mathbf{E}_{tot}(\mathbf{r},\omega) = \frac{\mathbf{E}_{in}(\mathbf{r},\omega)}{1-\frac{\omega^2}{c^2}\mathbf{G}_B(\mathbf{r},\mathbf{r}',\omega)\Delta\varepsilon(\omega)V_{QE}} \tag{A2}$$

For the quantum optical problem, we describe the system using the multiple-scattering approach for atoms in the form of point-like harmonic oscillators developed in Ref. [4], and consider the case where the single QE is initially in the ground state. In this context, we note that the formulation in Ref. [4] is performed in terms of specialized functions $\mathbf{F}(\mathbf{r},\omega)$ and $\mathbf{K}(\mathbf{r},\mathbf{r}',\omega)$, which differ from the electric field $\mathbf{E}(\mathbf{r},\omega)$ and the Green tensor $\mathbf{G}(\mathbf{r},\mathbf{r}',\omega)$ only when the two position arguments coincide at the position of one of the point scatterers. This construction effectively recovers the correct sum rule when integrating the electric field across the point scatterer, despite the fact that the Green tensor is known to diverge in this limit. In the present approach, for which we compare to the result of integration over the finite volume of the QE, we do not make this distinction and can therefore simply set $\mathbf{F}(\mathbf{r},\omega) = \mathbf{E}(\mathbf{r},\omega)$ and $\mathbf{K}(\mathbf{r},\mathbf{r}',\omega) = \mathbf{G}(\mathbf{r},\mathbf{r}',\omega)$. Otherwise following the approach of Ref [4], we can express the total electric field operator at the position of the QE as

$$\hat{\mathbf{E}}_{\text{tot}}(\mathbf{r},\omega) = \frac{\hat{\mathbf{E}}_{in}(\mathbf{r},\omega)}{1-\mathbf{G}_B(\mathbf{r},\mathbf{r}',\omega)V(\omega)} \tag{A3}$$

in which $\hat{\mathbf{E}}_{in}$ is the electric field operator of the incoming field, and $V(\omega)$ is the scattering potential produced by the QE and is given by [4]:

$$V(\omega) = \left(\frac{\mu_{QE}^2 \omega^2}{\hbar\varepsilon_0 c^2}\right)\left(\frac{2\omega_{QE}}{\omega^2-\omega_{QE}^2}\right) \tag{A4}$$

Comparing Eqs. (A2) and (A3), we can identify the connection between the permittivity and the dipole moment as

$$\Delta\varepsilon(\omega)V_{QE} = \frac{\mu_{QE}^2}{\hbar\varepsilon_0}\frac{2\omega_{QE}}{\omega^2-\omega_{QE}^2} \tag{A5}$$

and by rewriting slightly and including a non-radiative decay rate for the QE, we can express this in the form of the Lorentz oscillator model for the permittivity as

$$\Delta\varepsilon(\omega) = \frac{\mu_{QE}^2}{\hbar\varepsilon_0 V_{QE}}\left(\frac{2\omega_{QE}}{\omega_{QE}^2-\omega^2-2i\gamma_{QE}\omega}\right) \tag{A6}$$

Finally, comparing to Eq. (3) in the main text, we find that the dipole moment of the QE can be expressed in terms of the oscillator strength in the exact form of Eq. (14) in the main text.

Using the parameters of the QE in this work, $f=7\times 10^{-3}$, 20 nm radius and $\omega_{QE} = 1213.66\ 10^{12}\,\text{rad s}^{-1}$, we find $\mu_{QE} = 3.645\times 10^{-28}$ [Col·m], which is consistent with typical values for colloidal quantum dots [5-7].